# Toward a Theory on the Stability of Protein Folding: Challenges for Folding Models

Walter Simmons

Department of Physics and Astronomy

University of Hawaii at Manoa

Honolulu, HI 96822

Joel L. Weiner

Department of Mathematics

University of Hawaii at Manoa

Honolulu, HI 96822

# Abstract

We adopt the point of view that analysis of the stability of the protein folding process is central to understanding the underlying physics of folding. Stability of the folding process means that many perturbations do not disrupt the progress from the random coil to the native state. In this paper we explore the stability of folding using established methods from physics and mathematics. Our result is a preliminary theory of the physics of folding. We suggest some tests of these ideas using folding simulations.

We begin by supposing that folding events are related in some way to mechanical waves on the molecule. We adopt an analytical approach to the physics which was pioneered by M.V. Berry, (in another context), based upon mathematics developed mainly by R. Thom and V.I. Arnold. We find that the stability of the folding process can be understood in terms of structures known as caustics, which occur in many kinds of wave phenomena.

The picture that emerges is that natural selection has given us a set of protein molecules which have mechanical waves that propagate according to several mathematically specific restrictions.

Successful simulations of folding can be used to test and constrain these wave motions.

With some additional assumptions the theory explains or is consistent with a number of experimental facts about folding.

We emphasize that this wave-based approach is fundamentally different from energy-based approaches.

# Contents

Introduction

Rogue ocean waves that appear randomly and sometimes sink ships, bands of light seen on the bottom of a pool or stream on a sunny day, certain seismic events, and some recently discovered electronic effects in graphene, are highly diverse phenomena in the purview of the fields of fluid dynamics, optics, acoustics, and quantum mechanics but all have many aspects in common. (1) (2) (3) (4) (5) (6)

These phenomena are forms of caustics, which arise in wave propagation, wherein the waves pile up to very high amplitude, usually along a curve. Caustics, in contrast to other wave forms, arise only when certain specific conditions apply to the phases of the waves. (7) (8) (9) (10)

Broadly speaking, caustics can be divided into two categories: structurally stable versus unstable caustics. By structural stability, we mean that the general shape of the caustic is not sensitive to small details of the propagation process. Unstable caustics change form radically when the details of formation are modified even slightly. An example of an unstable caustic is a point image on a screen formed using a thin lens; any small defect in the lens blurs the image from its ideal point shape. Examples of stable caustics are the parallel bands of light that sometimes appear and persist on the bottom of a swimming pool on a sunny day; their form is relatively independent of the details of the surface motion of the water.

The study of the motion of waves is usually developed in terms of equations of motion and their solutions. In many cases this analysis approach is not very practical. Another approach involves treating waves as geometrical objects of study and identifying special features that serve to define the geometry of the waves. This approach is particularly well suited to the study of the stability of processes involving waves.

Following upon well-known developments in mathematics by R. Thom, V.I. Arnold, E.C. Zeeman, and others, M.V. Berry pioneered the application of this geometric approach to stable optical caustics; this is a *quantitative* theory. These bright curves and surfaces can be quite complicated but they are formed from a small set of stable, geometrically distinct shapes and the theory and experiment are in excellent, wide-ranging, agreement.

In a typical experimental arrangement, light from a single source is scattered in all directions from some set of lenses and/or mirrors. A caustic is a very bright curve that appears on a nearby screen. If the caustic is structurally stable, then minor adjustments to the shapes of the lenses and mirrors do not significantly change the shape of the curve on the screen.

As we have said, these structurally stable regions of high amplitude occur in optics, fluid dynamics, and acoustics, as well as other fields and, in some cases, quantitative theories have been developed; in other cases, especially outside of the physical sciences and engineering, *qualitative* theories have been presented by many authors; unfortunately, some of the qualitative theories have attracted criticism.

Motivated by the appearance of waves of variable wave-speed in an analytic model of protein folding, we suggested (11) that a quantitative theory of some parts of the folding process might be developed using the mathematical structure that underlies the physics of stable optical caustics.

Besides important empirical observations about folding, a key reason for pursuing this approach is that the mathematics and physics places very strong restrictions upon wave behavior at caustics. If this same mathematics and physics is involved in folding, then it will dominate certain folding processes and is deterministic.

Rogue water waves make a particularly interesting parallel with folding. These large waves appear suddenly and in random places. They arise because the wave-speed of gravity waves is frequency dependent in deep water. The faster waves can catch up with the slower waves leading to a sudden caustic. In a protein, we conjecture, features defined by the amino acid sequence pick out a region of the molecule for large amplitude motion.

For example, when viewed from a stability perspective, mutations of a stable protein have either no impact upon stability, change the protein to another stable form, or make the protein unstable (and hence probably not present in the data bases of naturally occurring proteins). The mathematics and physics of these kinds of transitions are available to characterize both stability and structural change.

In this paper, we re-introduce the ideas presented earlier and amplify upon them in detail. Our intended readership is scientists who model proteins and those who study mutations and protein structure. We will suggest some ways that the theoretical ideas discussed here might be explored in simulations.

## Background Biology

The process of protein manufacture in simple cells accounts for about 75% of the cell's power utilization. A typical simple cell manufactures several thousand proteins (12). The process of translating the information encoded in the DNA into folded proteins represents the principle flow of genetic information in simple cells and is essential in all cellular life.

The proteins are linear chains of amino acids, (twenty in number), which are formed and extruded into the intra-cellular fluid where they fold.

The work of Anfinsen showed that unfolded proteins of length up to about 350 residues fold spontaneously in water. This demonstrates that (1) the information needed to describe the fold is in the sequence, (2) the mechanism of folding is present in the molecule (plus water) itself, and (3) the details of the initial condition of the random coil are not determinant of the final form of the native state.

The random coil and the folded protein are often separated thermodynamically by only a small difference in the Gibbs potential. Within this short energy separation lies a phase transition. Aside from the energy needed to build the molecule and place it in solution, little additional energy is required to fold it.

A common pattern is that the entropy changes rapidly at the inception of folding. This is consistent with the idea that long segments of the chain are pinned by intra-chain contacts. The enthalpy changes later consistent with the formation of many bonds and detailed rearrangements.

The most striking feature of protein folding is the unique nature of the native state and the insensitivity of the folded form to perturbations during the folding process, including initial shapes of the random coil. In fact, this is one of the most puzzling phenomena in science (13) (14) (15) (16) (17) (18) (19) (20) (21) (22) (23).

One can easily construct a lengthy list of reasons why proteins should not fold to unique forms in water solution.  The lack of significant energy, interaction with the water, and the vast number of possible conformations that must be sorted out are three prominent issues.  Others include the list of small forces that are at play locally and globally, and the fact that generally speaking the rates of chemical processes are temperature sensitive, implying that parallel steps may complete in different order depending upon temperature.

Nevertheless, proteins do fold to their unique final states.  The best explanations as to why all these difficulties do not block reliable folding usually relate in some way to natural selection and the potential energy landscape; biological molecules have been selected to fold reliably.

The folding process terminates with the formation of chemical bonds; often this is an annealing process.

The current state of the art in modeling entails fitting empirical parameters using known folds, applying classical statistical mechanics or molecular dynamics, and usually making some simplifying assumptions about the potential energy surface. For many specific molecules this approach is quite successful. The reverse process, protein engineering, is more straight-forward and has been highly successful with libraries of elementary structures.

In a 2011 essay in Nature Structural and Molecular Biology, Stephen C. Harrison wrote (24),

 "Can we carry out, conceptually or computationally, the fundamental transformation from one dimension of information to three that is embodied in the central dogma? Can we robustly predict how a protein will fold? Can we deduce structure and function from sequence? There are clear hints that these questions will ultimately have positive answers, but we are no more confident of a timeline for the answers than we were in 1971."

## Properties of Stable Optical Caustics

In thinking about optical caustics, it may be useful to look at the beautiful photographs of caustics in the papers of M.V. Berry (3) (7). Wikipedia has some photos and some diagrams showing the relationships between wave-fronts and caustics. We also suggest examination of the electronic caustics in Cheianov, et al, (6) figures 3B, 3C, and 3D. Those illustrate the stability of a caustic in the presence of perturbations.

The caustics are typically three-dimensional but are usually viewed on a screen or table top. Throughout this paper, when discussing optical caustics, we assume the geometric optics limit in which the wavelength is small compared to other length scales. (At wavelength scale in optics or quantum mechanics the caustics are decorated with beautiful diffraction patterns).

An essential property of stable caustics in physics is that, beyond the energy required to set waves in motion, no additional energy is required to form the caustics. The formation process is geometrical and is fully determined in optics by the lenses and mirrors.

If the shape of the wave-front emerging from the scattering region is regarded as an initial value specification for the formation of the caustic down field, then the shape of the caustic is not significantly altered by small changes in the initial wave-front. Neither is it altered significantly by small changes along the path of propagation, such as the index of refraction.

As will be discussed in another section, this insensitivity follows from the underlying mathematics.

According to the mathematics, the number of elementary stable caustics is fixed; in fact the number is seven. More complicated stable caustics can form when multiple elementary caustics appear on the same screen. It is common for the caustics to form a network structure of elementary caustics; the ribbon-like bright lines on the bottom of a swimming pool are often cited as an example of this situation.

On the subject of the relevant physics and mathematics, Fermat's principle is the starting point: the time from the source to a point on the screen is minimized (technically, extremized). The quantitative theory of optical caustics is based upon singularity theory, of which, Thom's Catastrophe Theory is a sub-set. Note that this is a rather different application of Catastrophe Theory than that which is often used in potential theory.

Finally, we note that the propagation of the light stops when it encounters the image plane of a camera or the eye. This is absorption and differs from the scattering (refraction and/or reflection) that creates the caustic.

## The Analogy

We summarize here, in broad terms, the analogy between stable caustics and protein folds.

**Energy:**

> Caustic formation requires no additional energy beyond that required to power the light source. Energy considerations are not deterministic of the structure.
>
> Protein folding requires little energy and, in fact, energy may not be the most important physical quantity in folding.

**Insensitivity to Perturbations:**

> Caustics can be stable against perturbations, including initial conditions.
>
> The native state of a protein is stable against perturbations in the initial random coil form and against perturbations during folding. These perturbations can be directly in the shape, such as initial random coil shape variations, or in the mechanism of folding such as small pH changes.
>
> The native state is not significantly changed by temperature variations (over some limited range) and hence thermal agitation does not disrupt the folded form.

## Unique Final Form:

Proteins fold to form unique final structures. Stable caustics are unique in shape up to a coordinate transformation.

Elementary stable caustics are only seven in number. More complicated multi-caustic networks are known to form. The more complicated catastrophes unfold into simpler catastrophes. For example, a cusp looks like two fold caustics with a point in common.

The folding of proteins to unique final structures is essential to the biology.

Much research (25) (26) (27) (28) (29) supports the idea that complicated structures are built up from simpler ones. For example, domains built up from helixes, beta sheets, and coil.

## Insensitivity to Initial Conditions

Caustics are insensitive to small perturbations in the initial wave form.

The random coil has a large number of initial conformations. All of these progress to the unique native state.

## Global Action

A singularity in caustic theory arises from an integration of wave motion over a surface or volume.

Folding involves global and local sequence structure.

**Termination of Formation:**

> Caustics formation terminates with an interaction with a screen.
>
> Protein folding terminates when bonds form to lock in the final state.

**Wave Phenomenon:**

> Caustics originate in wave motion but may also occur in shock waves. Light waves propagate independently of one another but end up at the same space-time locations.
>
> Protein folding may be related to torsion waves of variable speed. The proteins of an ensemble fold independently of one another but end up with the same shape.

**Changes and Mutations**

> When the mirrors and/or lenses are modified in the presence of a caustic, the caustic is stable against small changes but eventually it will undergo a major change. It may become unstable or change into another caustic depending, in a predictable way, upon the details.
>
> More specifically, a change in the dimensionality of the physical arrangement makes one kind of caustic impossible but may permit another (see below for more detail).
>
> Mutations either make no change in structure, lead to an unstable structure, or change to a different structure.

## State Variables and Control Parameters in Wave Motion

There are two broad classes of variables of interest in wave motion: control parameters and state variables. It is important to distinguish these classes. For the purposes of this descriptive section, we do not require that these technical terms be defined carefully.

Control parameters are usually physical quantities that an experimenter can change. An example is the position and orientation of the screen upon which the caustic is viewed. Caustics are viewed in the space of all relevant control parameters.

For light, state variables directly affect the time of travel along various possible paths. An example is the index of refraction along the possible light paths. According to Fermat's principle, the light will follow the path defined by the state variables that takes the least time. In more technical terms, Fermat requires that the optical path of a ray be stationary with respect to variation of paths between points represented by state variables and control parameters.

As we shall note, below, stable caustics occur only if the number of control parameters is 5 or fewer. There can be many state variables.

## Toward a New Theory

If natural selection has resulted in a set of biological proteins that have a special property, derivable from the sequence, that causes them to fold to unique structures in the presence of potentially disrupting influences, then what is the physical nature of that 'special property'?

From the perspective of the physics and mathematics we have suggested that folding entails the propagation of mechanical waves, presumably mainly torsion waves, which have variable wave-speed. The special property we are looking for is a set of constraints on the phase (or time-delay function) of these waves.

If the theory described here agrees with observations then it has the potential to become a completely quantitative theory. When completed, it might potentially relate the geometry of mechanical waves to the chemical and physical properties of the residues.

Another advantage is that the mathematics has essentially all been worked out. What remains to test and complete the theory is discussed in the next section.

In this section, we discuss some of the aspects of this problem that can be inferred using available information.

We shall begin by supposing the existence of some, perhaps multi-dimensional, mathematical variable that is derivable from the sequence over some length of the molecule and which effects the formation of a caustic at another locus. For simplicity, we start by using a scalar variable $X$; (more commonly, this actually stands for a collection of state variables). The control parameters are similarly symbolized by $\xi$.

We denote the phase function by $\Phi(X,\xi)$.

The two essential conditions for the appearance of a caustic at a specific point are,

$$\frac{\partial}{\partial X}\Phi(X,\xi)=0$$

and

$$\frac{\partial^2}{\partial X^2}\Phi(X,\xi)=0$$

These conditions determine points in both variables but Thom's theory applies around this point, thereby making it enormously powerful in physics, especially with respect to issues of stability. The equation of the caustic, expressed in the control variables $\xi$, is obtained by eliminating the variable $X$ in the two equations. It is not necessary to go into how to carry that out, as all the cases have already been worked out and catalogued. Moreover, there are only seven solutions for stable caustics and they arise from simple polynomials in the state variables $(X)$ and control parameters $(\xi)$. Any mathematical form, (up to a change in coordinates), for the phase that is not on Thom's list is not stable.

So far, our theory looks like this: natural selection picked out molecules that have mechanical waves whose phases depend upon state variables and control parameters. The phases, (or propagation delay functions), are determined by some property involving the residues and derivable from the sequence. The shape of the resulting caustic is expressed in control variables, $\xi$ and this determines the shape of the protein. The possible shapes are limited to seven possibilities. For stable caustics the maximum number of control variables is five. An analogy with the bands seen on a stream bed is appropriate. The dimensions of the problem pick out one catastrophe; the details of the surface motion are not important.

We called this property, $X$, in the interest of simplicity. $X$ might be analogous to an index of refraction which is dependent upon position. It might be a simple scalar derivable from, say, the polarity of the residues. However, it is likely to be more complicated.

At present, we have only limited information about the waves of interest. They must, of course propagate in a space of three spatial dimensions and three dimensions of orientation at each point. We do not know that all of these dimensions are involved in picking out a caustic; the $(\Phi, \Psi)$ angles and the distance along the chain may be sufficient.

If the waves and their phases can be identified in simulations, then mutations involving substitutions can potentially identify the precise dependence of the phases upon the structure of the residues.

Even in the simple form just described the theory has a number of interesting features. The fact that the native state is not sensitive to small variations in the initial conditions of the random coil is important. Similarly, at least qualitatively speaking, the stability of the process of the formation of native state against perturbations also follows by the same argument.

All of the features described in the Analog section, above, are described at either quantitatively or qualitatively by this theory.

We offer now some speculations to expand the theory.

We have not established a specific relationship between the shape of the caustic and the shape of the molecule. However, it seems reasonable that the higher order of the singularity, the more complicated the associated shape.

We assume that only certain aspects of the molecule are associated with these caustics.

Natural conjectures would be that for a fold caustic, the shape is a hairpin, for a swallowtail catastrophe there is a crossing point and two sharp bends so this may form the initial configuration of a helix, where the first three intra-chain contacts are needed to get the helix formation process started.

Based upon the appearance of high-degree polynomials in the state variables ($X$, above), corresponding to the appearance of caustics, we have suggested that heterogeneity of the sequence is critical in producing the formation of a caustic and hence of forming a fold.  This, in turn, hints that there is a significant scale-length, over which the sequence is heterogeneous, and hence suggests that homopolymers and relatively smooth random polymers do not reliably fold in this theory. This also follows from the fact that we are only considering waves of variable speed. (In our analytical models, variable speeds will not arise in homogeneous situations).

## Challenges

The most direct way to test and limit this theory is to study the propagation of waves during folding.  We need to know how many dimensions are relevant, how the phase is changed by the physical and chemical properties of the residues (polarity, moment of inertia, nearest neighbor interactions, etc.), the wavelengths, and whether traveling-waves and/or solitons are involved.

Another approach, assuming the waves can be defined in the context of various models, is to find out if the phase conditions, above, are, in fact, satisfied and if the molecule assumes the shape of the caustic around such points.

## Conclusions

We have presented our analysis of the stability of the protein folding process in terms of established physics and mathematics and found a strong correspondence between caustics and protein folds. We restated our findings as a simple statement of a theory of folding.

We propose that torsion waves of variable wave-speed conspire to form caustics in a way similar to the way in which gravity waves of variable wave-speed conspire to form caustics. We suggest that natural selection has left us with molecules that have this unusual but well defined property.

It seems unlikely that waves of such potentially complicated form will ever be understood analytically. We have therefore emphasized a geometric approach involving caustics. Technically speaking, the wave is treated as a geometric object and the study of its singularities is a quantitative approach to understanding the wave behavior.

If a form of this theory is correct, then important aspects of folding can be *quantitatively* determined from data on the phases of mechanical waves on the molecule. In particular, the key data elements would be the manner in which the phase is changed by the interaction of the wave with the residues

The most fundamental way to test and expand this theory is to find the behavior described in equations in the Theory section, above, in waves of variable wave-speed. This is a formidable problem in view of the large number of forces and moving components that come into play.

Simulations of folding can be used to explore the ideas presented here by looking for the patterns we have described: waves of variable wave-speed, heterogeneity of the sequence when an amino acid substitution results in a new fold, decomposition of the form of the molecule just prior to the formation of bonds or of geometrical hindrance, and the study of nearly homogeneous and random sequences to determine the length scales involved.

## Mathematics and Physics

In these technical sections we discuss material somewhat more advanced than that used in the main part of the paper.

I.) Mathematics Literature:

A wide range of sources are available for the study of the mathematics. For completeness, we present here a brief literature survey.

The book by P.T. Saunders is an excellent introduction of the subject of catastrophe theory. He has only short section on optical caustics.

The book by R. Gilmore is comprehensive and has a clear section on caustics.

The best sources on optical caustics are the papers by Berry and his colleagues. In particular, the review of catastrophe optics by Berry and Upstill is excellent. Various other papers by Berry and colleagues are also excellent.

II.) Caustics:

In this section, we present a brief guide to the technical treatment of the theory of optical caustics.

(Parenthetically, we comment that the treatment requires the use of vector calculus of multiple variables but we chose our example so that elementary calculus of one variable could be be used.)

The set-up is a standard text-book optical arrangement: a source of light illuminates an object. Light from every point on the object reaches each point on an image screen located some distance away. The coordinates of points on the screen are control variables and are described here by some equations in variables described by the Greek symbol, $\xi$. The state variables, the variables such as local path and velocity that appear in the action are written in Latin (x).

The light has a phase, or time-delay function, $\Phi(X, \xi_1, \xi_2)$, which depends upon both control parameters and state variables. This function carries all the information that we require about the object and about where it is viewed.

The electric field at the image is given by the Fresnel integral, which we simplify to

$$E(\xi_1, \xi_2) = \int dX \, \exp(i\Phi(X, \xi_1, \xi_2))$$

A key point is that such integrals get their maximum contribution from those points where the phase function is stationary; that is to say, the gradient of the phase function is zero.

We begin by stepping inside the integral sign. We shall assume that for the particular set-up we that have, a stable caustic exists. This would be easily identified as a very bright area on the image. (Unstable caustics will be discussed below).

The first step in understanding the stable caustic is to set the gradient of the phase with respect to the Latin variables (state variables) to zero and also to set the next higher derivative with respect to these variables to zero. (Please see the Theory section, above). The two resulting equations pick out a set of points $(\xi_1, \xi_2)$ that correspond to the singularity.

Next, we apply Thom's powerful theorems. From this, we learn that the phase function must have one of seven possible forms. In order to make this more specific, let's write down one possibility (the cusp catastrophe):

$$\Phi(X, \xi_1, \xi_2) = X^4 + \xi_1 X^2 + \xi_2 X$$

The calculation steps can be constructed as follows. First is the condition, following from Fermat's principle, that the phase (time delay) is at a minimum (or extremum)

$$\frac{d}{dX}\Phi(X, \xi_1, \xi_2) = 4X^3 + 2\xi_1 X + \xi_2 \equiv 0$$

Next, the condition for the appearance of the caustic,

$$\frac{d^2}{dX^2}\Phi(X, \xi_1, \xi_2) = 12X^2 + 2\xi_1 X \equiv 0$$

The final step is to find the places on the screen, i.e. values of $(\xi_1, \xi_2)$, at which the phase is stationary and the second derivatives also vanish.

Since both of the above equations must be simultaneously true, this can be done algebraically by eliminating $X$ from the above two equations with the result,

$$8\xi_1^3 + 27\xi_2^2 = 0$$

That is the equation describing the caustic as seen on the screen. As is well known, this is a cusp.

As an example of stability (8), let us examine what happens to the cusp caustic, above, which arises from the variable $X$, for a perturbation in another variable, $Y$, of the form $\xi_1 Y^2 + \xi_2 Y$. We calculate, first the condition from Fermat's principle,

$$\frac{\partial}{\partial Y}\Phi(X,Y,\xi_1,\xi_2) = 2\xi_1 Y + \xi_2 = 0$$

and second the condition for the appearance of a caustic in this variable.

$$\frac{\partial^2}{\partial Y^2}\Phi(X,Y,\xi_1,\xi_2) = 2\xi_1 = 0$$

There is no impact upon the caustic due to this other variable at an arbitrary point on the screen.

There are seven stable catastrophes and hence seven stable optical caustics. Other caustics are unstable; that is, any small change in the Greek variables significantly alters the form of the caustic.

As noted before, a point image on a screen is not one of the stable caustics; it is therefore unstable. When the phase depends upon more than five control parameters, it produces only unstable caustics. Technically, there are catastrophes that have a looser form of stability (homeomorphisms) but that is beyond the scope of this paper.

Caustics are classified by their co-dimension, which is the dimension of the space containing the caustic minus the dimension of the caustic itself. The result is often just one. (There is just one equation in addition to the condition implied by Fermat's theorem.). Berry remarks, "Loosely stated, [the codimension] is the minimum number of essential control parameters of a space that contains the singularity."

We emphasize, again, that in view of the fact that there are only seven caustics, the mathematics has been worked out rather completely and is available from the literature.

1. **White, B.S. & Fornberg, B.** On the Chance of Freak Waves at Sea. *Journal of Fluid Mechanics.* 1998, Vol. 355, 113.

2. **Brown, M.G.** Space-time surface gravity wave caustics. *Wave Motion.* 2001, Vol. 33, 117.

3. **Berry, M.V,& Upstill.** Catastrophe Optics. Morppologies of Caustics and Their Diffraction Patterns. *Progress in Optics.* 1980, Vol. 18, 257.

4. **Berry, M.V.** Singularities in waves and rays. [book auth.] R. ,Kleman, M., and Poirier, J.-P. Balian. *Physics of Defects.* Amsterdam : North Holland, 1981.

5. **Chapman, C.H. & Drummond, R.** Body-Wave Seismograms in Inhomogeneous Media Using Maslov Asymptotic Theory. *Bulletin of the Seismological Society of America.* 1987, Vol. 72, S277.

6. **Cheianov, V.V. Fal'ko, V., and Altshuler, B.L.** The Focusing of Electron flow and a Veselago Lens in Graphene p-n Junctions . *Science.* 2007, Vol. 315, 1252.

7. **Berry, M.V.** Beyond Rainbows. *Current Science .* 1990, Vol. 59, 1252.

8. —. Waves and Thom's Theorem. *Advances in Physics.* 1976, Vol. 25, 1.

9. **Gilmore, R.** *Catastrophe Theory for Scientists and Engineers.* New York : Wiley & Sons, 1981.

10. **Saunders, P.T.** *Catastrophe Theory.* Cambridge : Cambridge University Press, 1980.

11. *Physics of Caustics and Protein Folding: Mathematical Parallels.* **Simmons, W. & Weiner, J.L.** s.l. : arXiv 1108.2740 , 2011.

12. **Lane, N.** Energetics and genetics across the prokaryote-eukaryote divide. *Biology Direct.* 2011, Vol. 6, 35.

13. **Dagget, V. & Fersht,A.R.** Is there a unifying mechanism for protein folding. *TRENDS in Biochemical Sciences.* 2009, Vol. 28, 18.

14. **Dill, K.A., Ozkan, S.B., Shell, M.S., and Weikl, T.T.** The Protein Folding Problem. *Ann. Rev. Bioph.* 2008, Vol. 37, 289.

15. **Rose, G.D., Fleming, P.J., Banavar, J.R. and Maritan, A.** A backbone-based theory of protein folding. *PNAS.* 2006, Vol. 103, 16623.

16. **Onuchic, J.N. & Wolynes, P.G.** Theory of Protein Folding. *Current Opinion in Structural Biology.* 2004, Vol. 14, 70.

17. **Yang, J.S., Chen, W.W., Skolnick, J., and Shakhnovich, E.I.** All-Atom Ab Initio Folding of a Diverse Set of Proteins. *Structure.* 2007, Vol. 15, 53.

18. **Dodson, E.J.** Protein predictions. *Nature.* 2007, Vol. 450, 176.

19. **Wolynes, P.G.** Recent successes of teh energy landscape theory of protein folding and function. *Quarterly Reviews of Biophysics.* 2004, Vol. 38, 4.

20. **Shakhnovich, E.** Theoretical studies of protein-folding thermodynamikcs and kinetics. *Current Opinion in Structural Biology.* 1997, Vol. 7, 29.

21. **Wolynes, P.G. Onuchic, J.N. and Thirumalai, D.** Navigating the Folding Routes. *Science.* 1995, Vol. 267, 1619.

22. **Karplus, M. & Kurlyan, J.** Molecular dynamics and protein function. *PNAS.* 2005, Vol. 102, 6679.

23. **Hubner, I.A., Deeds, E.J., and Shakhnovich, E.I.** Understanding ensemble protein folding at atomic detail. *PNAS.* 2006, Vol. 17747.

24. **Harrison, S.C.** Forty years after. *Nature Structural & Molecular Biology.* 2011, Vol. 18, 1304.

25. **Koonin, E.V, Wolf, Y.I. and Karev, G.P.** The structure of the protein universe and genome evolution. *Nature.* 2002, Vol. 420, 218.

26. **Choi, I-G. & Kim, S-H.** Evolution of protein structural classes and protein sequence families. *PNAS.* 2006, Vol. 103, 14056.

27. **Chothia, C., Hubard, T.m Barns, H. and Murzin, A.** Protein folds in the all beta and all alpha classes. *Annu. Rev. Biophys. Biomol Structure .* 1997, Vol. 26, 597.

28. **Chothia, C. & Finkelstein, A. V.** The Classifiction and Origins of Protein Folding Patterns. *Annu. Rev. Biochem. .* 1990, Vol. 59, 1007.

29. **Bashton, M. & Chothia, C.** The Generation of New Protein Functions by the Combination of Domains. *Structure.* 2007, Vol. 15, 85.